\documentclass{IEEEtran}
\pdfoutput=1
\usepackage{amsmath}
\usepackage{cite}
\usepackage{color}
\usepackage{booktabs}
\usepackage{graphicx}
\usepackage{subfigure}

\renewcommand{\vec}[1]{{\bf #1}}
\newcommand{\sech}{{\mbox{sech}}}

\title{Modeling sparse connectivity between\\ underlying brain sources for EEG/MEG}
\author{Stefan~Haufe, Ryota~Tomioka, Guido~Nolte, Klaus-Robert~M\"uller and Motoaki~Kawanabe
\thanks{S.~Haufe and K.-R. M\"uller are with the Berlin Institute of Technology, Germany.}\thanks{R.~Tomioka is with the University of Tokyo, Japan.}\thanks{G.~Nolte and M.~Kawanabe are with Fraunhofer Institute FIRST, Berlin, Germany.}}

\date{\today}
\begin{document}
  \maketitle 
  
  \begin{abstract}
We propose a novel technique to assess functional brain connectivity
in EEG/MEG signals. Our method, called Sparsely-Connected Sources
Analysis (SCSA), can overcome the problem of volume conduction by
modeling neural data innovatively with the following ingredients: (a)
the EEG is assumed to be a linear mixture of correlated sources following a
multivariate autoregressive (MVAR) model, (b) the demixing is estimated jointly with the source MVAR parameters, 
(c) overfitting is avoided by using the
Group Lasso penalty. This approach allows to extract the appropriate
level cross-talk between the extracted sources and in this manner we
obtain a sparse data-driven model of functional connectivity. We
demonstrate the usefulness of SCSA with simulated data, and compare to
a number of existing algorithms with excellent results.
\end{abstract}
  
\section{Introduction}
  
\subsection{Functional brain connectivity}
The analysis of neural connectivity plays a crucial role for understanding the general functioning of the brain. In the past two decades such analysis has become possible thanks to tremendous progress that has been made in the fields of neuroimaging and mathematical modeling. Today, a multiplicity of imaging modalities exists, allowing to monitor brain dynamics at different spatial and temporal scales. 

Given multiple simultaneously-recorded time-series reflecting neural activity in different brain regions, a functional (task-related) connection (sometimes also called information flow or (causal) interaction in this paper) between two regions is commonly inferred, if a significant time-lagged influence between the corresponding time-series is found. Different measures have been proposed for quantifying this influence, most of them being formulated either in terms of the cross-spectrum (e.g., coherence, phase slope index \cite{NolZieNik08}) or an autoregressive models (e.g., Granger causality \cite{granger}, directed transfer function \cite{dtf}, partial directed coherence \cite{pdc}, \cite{HauNolMueKra08}). 

\subsection{Volume conduction problem in EEG and MEG}
In electroencephalography (EEG) and magnetoencephalography (MEG), sensors are placed outside the head and the problem of volume conduction arises. That is, rather than measuring activity of only one brain site, each sensor captures a linear superposition of signals from all over the brain. This mixing introduces instantaneous correlations in the data, which can cause traditional analyses to detect spurious connectivity \cite{imagcoh}.  

\subsection{Existing source connectivity analyses}
Only recently, methods have been brought up, which qualify for EEG/MEG connectivity analysis, since they account for volume conduction effects. These methods can roughly be divided as follows.

One type of methods aims at providing meaningful connectivity estimates between sensors. The idea here is, that only the real part of the cross-spectrum and related quantities is affected by instantaneous effects. Thus, by using only the imaginary part, many traditional coupling measures can be made robust against volume-conduction \cite{imagcoh, NolZieNik08}. 

Another group of methods attempts to invert the mixing process in order to apply standard measures to the obtained source estimates. These methods can be further divided into (i) source-localization approaches (where sources are obtained as solutions to the EEG/MEG inverse problem), (ii) methods using statistical assumptions, and (iii) combined methods. The first approach is pursued, for example, in \cite{pmid17894381, pmid15792902}. Methods in the second category can be appealing, since they avoid finding an explicit inversion of the physical forward model. Instead, both the sources and the (de-)mixing transformation are estimated. To make such decomposition unique, assumptions have to be formulated, the choice of which is not so straightforward. We will now briefly review some possibilities for such assumptions.

Principal component analysis (PCA) and independent component analysis (ICA) are the most prominent linear decomposition techniques for multivariate data. Unfortunately, these methods contradict either with the goal of EEG/MEG connectivity analysis (assumption of independent sources in ICA\footnote{Although, under some circumstances this approach can be justified \cite{icaconn}.}) or even with the physics underlying EEG/MEG generation (assumption of orthogonal loadings in PCA). Nevertheless, both concepts have been successfully used in more sophisticated ways to find meaningful EEG/MEG decompositions. 


For example, an interesting use of ICA is proposed in \cite{mvarica}. The authors of this paper do not assume independence of the source traces, but rather argue that this property holds for the residuals of an MVAR model if no instantaneous correlations in the data exist. Hence, in their MVARICA approach they apply ICA to the residuals of a sensor-space MVAR model.


In this work, we first propose a single-step procedure to estimate 
all parameters (i.e. the mixing matrix and MVAR coefficients) of 
the linear mixing model of MVAR sources \cite{mvarica} based on 
temporal-domain convolutive ICA, instead of the combination of MVAR parameter fitting and
demixing by instantaneous ICA. Furthermore, the approach enables us to 
integrate a sparsity assumption on brain connectivity, i.e. 
on interactions between {\em underlying brain sources} via the Group 
Lasso penalty. The additional sparsity prior can avoid overfitting 
in practical applications and yields more interpretable estimators of 
brain connectivity. We remark that it is hard to incorporate such 
sparsity priors in MVARICA, since MVAR is fit to the {\em sensor signals}
where interactions (i.e. MVAR coefficients) are not at all sparse due 
to the volume conduction.

The remainder of the paper is organized as follows. In Section \ref{sec:csa}, 
our procedure will be explained step by step.
The correlated source model assumed in this paper will be defined in 
\ref{sec:model_assumption}. The identification procedure called connected
sources analysis (CSA) based on the convolutive ICA will be introduced 
(\ref{sec:cICA}) and followed by its sparse version,
sparse connected sources analysis (SCSA) with the Group Lasso prior 
(\ref{sec:sparse_reg}). The relations of our methods with existing approaches
such as MVARICA \cite{mvarica} and CICAAR \cite{cicaar} will be
elucidated in detail (\ref{sec:relation}). Finally, the optimization 
algorithms for CSA and SCSA will be explained
(\ref{sec:optimization}). 
We implemented two versions for SCSA, one based on L-BFGS and the
other by EM algorithm which is slower, but numerically more stable. 
The next section \ref{sec:experiment} will provide our experimental 
results on simulated data sequences emulating
realistic EEG recordings. The plausibility of our correlated source 
model will be discussed with future research directions in the context 
of computational neuroscience (Section \ref{sec:discussion}), before 
the concluding remarks (Section \ref{sec:conclusion}).

\section{Connected Sources Analysis with Sparsity Prior}
\label{sec:csa}

\subsection{MVAR for modeling causal interactions}
Autoregressive (AR) models are frequently used to define directed ``Granger-causal'' relations between time-series. The original procedure by Granger involves the comparison of two models for predicting a time series $z_i$, containing either past values of $z_i$ and $z_j$, or $z_i$ only \cite{granger}. If involvement of $z_j$ leads to a lower prediction error, (Granger-causal) information flow from $z_j$ to $z_i$ is inferred. Since this may lead to spurious detection of causality if both $z_i$ and $z_j$ are driven by a common confounder $z_*$, it is advisable to include the set $\{z_1,\ldots, z_M\}\setminus \{z_i,z_j\}$ of all other observable time series in both models. 

It has been pointed out, that pairwise analysis can be replaced by fitting one multivariate autoregressive (MVAR) model to the whole dataset, and that Granger-causal inference can be performed based on the estimated MVAR model coefficients (e.g., \cite{Valdes0501, HauNolMueKra08}). Several connectivity measures are derived from the MVAR coefficents \cite{dtf, pdc}, 
but probably the following definition is most straightforward
from Granger's argument that the cause should always precede the effect.
We say that time series $z_i$ has a causal influence on time series
$z_j$ if the present and past of the combined time series  $z_i$ and
$z_j$ can better predict the future of  $z_j$ than the present and
past of  $z_j$ alone. In the bivariate case this is equivalent to
saying that for at least one  $p \in \{ 1, \hdots, P \}$, the
coefficient  $H^{(p)}_{ji}$ corresponding to the interaction between
$z_j$ and $z_i$ at the $p$th time-lag is nonzero (
significantly different from zero). In the multivariate case, Granger causality also includes indirect causes not contained in non-vanishing  $H^{(p)}_{ji}$. 

\subsection{Correlated sources model}
\label{sec:model_assumption}
In this paper we propose a method for demixing the EEG/MEG signal into
causally interacting sources. 
We start from the same model as in \cite{mvarica}:
the sensor measurement is assumed to be generated as a linear instantaneous mixture of sources, which follow an MVAR model
\begin{eqnarray}\label{eq:eegmodel}
\vec{x}(t)& = & M \vec{s}(t) \\
\vec{s}(t) & = & \sum_{p=1} ^P H^{(p)} \vec{s}({t-p}) + \boldsymbol{\varepsilon}(t) \;.
\end{eqnarray}
Here, $\vec{x}(t)$ is the EEG/MEG signal at time $t$, $M$ is a mixing matrix representing the volume conduction effect, $\vec{s}(t)$ is the demixed (source) signal.
The sources at time $t$ are modeled as a linear combination of their $P$ past values plus an innovation term $\boldsymbol{\varepsilon}(t)$, according to an MVAR model with coefficient matrices ${H^{(p)}}$. 
In the standard MVAR analysis, the innovation 
$\boldsymbol{\varepsilon}(t)$ is a temporally- and
spatially-uncorrelated Gaussian sequence. In contrast, we
assume here that it is {\it i.i.d.} in time and the components are
subject to non-Gaussian distributions in order to apply blind source
separation (BSS) techniques based on higher-order statistics \cite{mvarica,cicaar}.

For simplicity, we deal with the case that the numbers of sensors and
sources are equal and the mixing matrix $M$ is invertible. When there
exist less sources than sensors, the problem falls into the current
setting after being preprocessed by PCA \cite{mvarica}. Under our model
assumptions, the innovation sequence can be obtained by a finite impulse response (FIR) filtering of the observation, i.e.
\begin{eqnarray}\label{eq:innmodel}
\boldsymbol{\varepsilon}(t) & = & M^{-1} \vec{x}(t) - \sum_{p=1} ^P H^{(p)} M^{-1} \vec{x}({t-p})\\
 & = & \sum_{p=0} ^P W^{(p)} \vec{x}({t-p}) \;,
\end{eqnarray}
where the filter coefficients are determined by the mixing matrix $M$
and the MVAR parameters $\{ H^{(p)} \}$ as
\begin{equation} \label{eq:vartrans}
W^{(p)} = \left\{
\begin{array}{ll}
M^{-1} & p = 0\\
-H^{(p)}M^{-1} & p > 0
\end{array}
\right. \;.
\end{equation}
Thanks to the non-Gaussianity assumption on the innovation 
${\boldsymbol \varepsilon}(t)$, we can use BSS techniques based on
higher-order statistics to identify the inverse filter $\{ W^{(p)}
\}$.
Since we would like to impose sparse connectivity as a plausible
prior information later on, it is preferable to apply temporal-domain
convolutive ICA algorithms. The obtained FIR coefficients 
$\{ W^{(p)} \}$ directly identify the mixing matrix $M$ and the MVAR
model of the same order $P$.

\subsection{Identification by convolutive ICA}
\label{sec:cICA}
We use temporal-domain convolutive ICA for inferring volume conduction
effects and causal interactions between extracted brain signals.
The model parameters can be identified based on the mild assumptions that the innovations are non-Gaussian and (spatially and temporally) independent. For EEG and MEG data, a super-Gaussian is prefered to a sub-Gaussian distribution, assuming that ongoing activity of brain networks is triggered by spontaneous local bursts. We here adopt the super-Gaussian $\sech$-distribution that was proposed in \cite{cicaar}. The Likelihood of the data under the model is then
\begin{eqnarray}\label{eq:likelihood}
\lefteqn{p(\{ {\bf x}(t) \}_{t=P+1}^T | \{ W^{(p)} \})} \nonumber \\
& = & |W^{(0)}|^{T-P} \prod_{t=P+1}^{T} \prod_{d = 1}^D \frac{1}{\pi} \sech \left( \varepsilon_d(t) \right) \;,
\end{eqnarray}
where ${\bf \varepsilon}(t) = M^{-1} {\bf x}(t) - \sum_{p=1} ^P
H^{(p)} M^{-1} \vec{x}({t-p})$.
The cost function to be minimized is the negative log-Likelihood 
%
%
\begin{eqnarray}\label{eq:negloglikelihood}
\lefteqn{{\cal L}(\{W^{(p)}\}) \ = \ (P-T) \log |W^{(0)}|} \nonumber
\\
& & \hspace*{1cm} - \sum_{t = P+1}^T \sum_{d=1}^D \log \left( \frac{1}{\pi} \sech\left(\varepsilon_d(t) \right) \right) \;.
\end{eqnarray}
%
The solution of Eq.~(\eqref{eq:negloglikelihood}) leads to the
estimators of the mixing matrix $M$ and the MVAR coefficients 
$\{ H^{(p)}\}$ via Eq.~(\eqref{eq:vartrans}). We will call this
procedure Connected Sources Analysis (CSA).

We remark that the temporal-domain algorithm of convolutive ICA has obvious 
indeterminacy due to permutations and sign flips. However, once we fix
a rule to chose one from all candidates, the cost function can be
considered as convex.

\subsection{Sparse connectivity as regularization}
\label{sec:sparse_reg}
In practice, we usually have to consider a long-range lag $P$ to explain
temporal structures of data sequences. However, this causes too many 
parameters to be estimated reliably.
Maximum-Likelihood estimation may easily lead to overfitting, especially if $T$ is small. For this reason, it is advisable to adopt a regularization scheme. Several authors have suggested that the complexity of MVAR models can be reduced by shrinking MVAR coefficients towards zero. In \cite{Valdes0501} and \cite{Sanchez0801}, MVAR-based functional brain connectivity is estimated from functional magnetic resonance imaging (fMRI) recordings using an $\ell_1$-norm based (Lasso) penalty, which has the property of shrinking some coefficients exactly to zero. In \cite{HauNolMueKra08} it is pointed out, that, by using a so-called Group Lasso penalty, whole connections between time-series can be pruned at once. In this approach, all coefficients $H^{(p)}_{ij}, p = 1, \hdots, P$ modeling the information flow from $\vec{s}_i$ to $\vec{s}_j$ are grouped together and can only be pruned jointly. From the practical standpoint such sparsification is very appealing, since fewer connections are much easier to interpret. But assuming sparse connectivity in fMRI data might also be justified from a neurophysiological point of view, since under appropriate experimental conditions only a few macroscopic brain areas are expected to show significant interaction. This reasoning also applies to EEG and MEG data.

We note that, besides the penalty-based approach, other strategies for obtaining sparse connectivity graphs exist. For example, post-hoc sparsification can be achieved for dense estimators by means of statistical testing \cite{kernelgranger2, HauNolMueKra08}. However, due to the compelling built-in regularization, we here adopt Group Lasso sparsification.

Before applying our regularization to the cost function of the correlated sources model, it is important to note that the sparsity assumption is only reasonable for the MVAR coefficients $\{H^{(p)} \}$, but not for the $W^{(p)}$ matrices which combine MVAR coefficients and the instantaneous demixing. Hence, in order to apply sparsifying regularization, one has to split the parameters into demixing and MVAR parts again, as in the original model Eq.~(\eqref{eq:eegmodel}). Since the offdiagonal elements $\{H^{(p)} \}$ correspond to interaction between sources, we propose to put a Group Lasso penalty on them analogously to \cite{HauNolMueKra08}. I.e., we penalize the sum of the $\ell_2$-norms of each of the groups $\{H_{df}^{(p)} \}, d \neq f$. 

Let $B := M^{-1}(=W^{(0)})$, $\vec{s}(t) = B \vec{x}(t)$ and
$\tilde{\vec{s}}(t) = \sum_{p=1}^P H^{(p)} \vec{s}({t-p})$.
The regularized cost function is
\begin{eqnarray}\label{eq:regnegloglikelihood}
\lefteqn{{\cal L^{\mbox{\tiny SCSA}}}(B, \{H^{(p)}\})} \nonumber\\
\nonumber  & = & (P-T) \log |B| + \lambda \sum\limits_{d \neq f} \left\| \left( {H_{df}^{(1)}}, \hdots, {H_{df}^{(P)}} \right)^\top \right\|_2\\
& & - \sum\limits_{t = P+1}^T \sum\limits_{d=1}^D \log \left(
\frac{1}{\pi} \sech\left( {\rm s}_d(t) 
- \tilde{\rm s}_d(t) 
\right) \right) \;,
\end{eqnarray}
$\lambda$ being a positive constant. The solution to Eq. (\eqref{eq:regnegloglikelihood}) 
for a choice of $\lambda$ is called
the Sparsely-Connected Sources Analysis (SCSA) estimate.

\subsection{Relation to other methods}
\label{sec:relation}
The proposed method extends previously suggested MVAR-based sparse
causal discovery approaches \cite{HauNolMueKra08, Valdes0501} by a
linear demixing, which is appropriate for EEG/MEG connectivity
analysis. Although the correlated sources model Eq.~\eqref{eq:eegmodel}
leads to an MVAR model of the observation sequence \cite{mvarica}, 
sparsity of the coefficients cannot be expected after mixing by volume
conduction effects.
Our method compares with MVARICA \cite{mvarica}, which uses the same model Eq.~\eqref{eq:eegmodel}, but estimates its parameters differently. More precisely, the authors of MVARICA suggest to first fit an MVAR model in sensor-space. The demixing can then be obtained by performing instantaneous ICA on the MVAR innovations, i.e., a dedicated contrast function (Infomax) is used to model independence of the innovations. The obtained sources follow an MVAR model with time-lagged effects (interactions), but ideally no instantaneous correlations (as caused by volume conduction). 

\begin{figure}[htb]
\begin{center}
\begin{tabular}{ccc}
\includegraphics[width=0.17\textwidth]{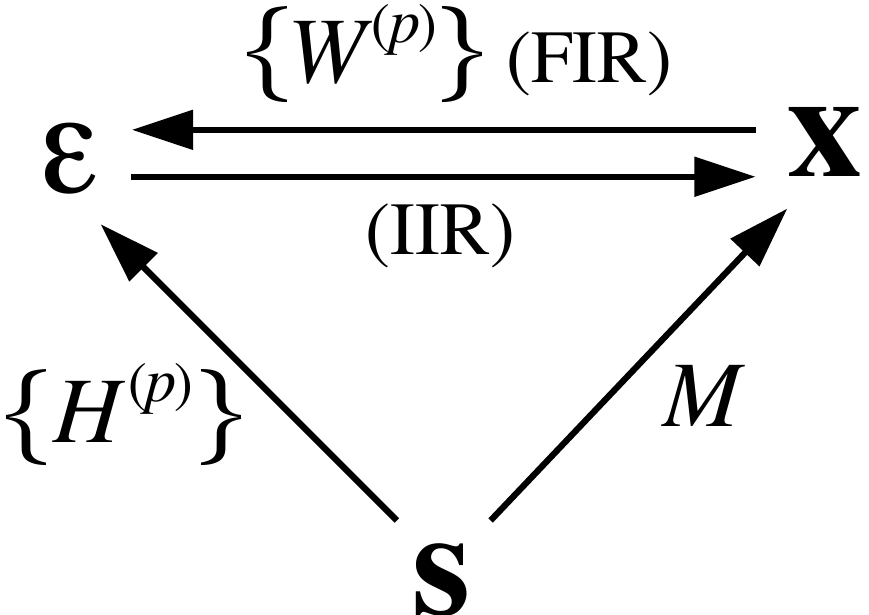} & \qquad &
\includegraphics[width=0.21\textwidth]{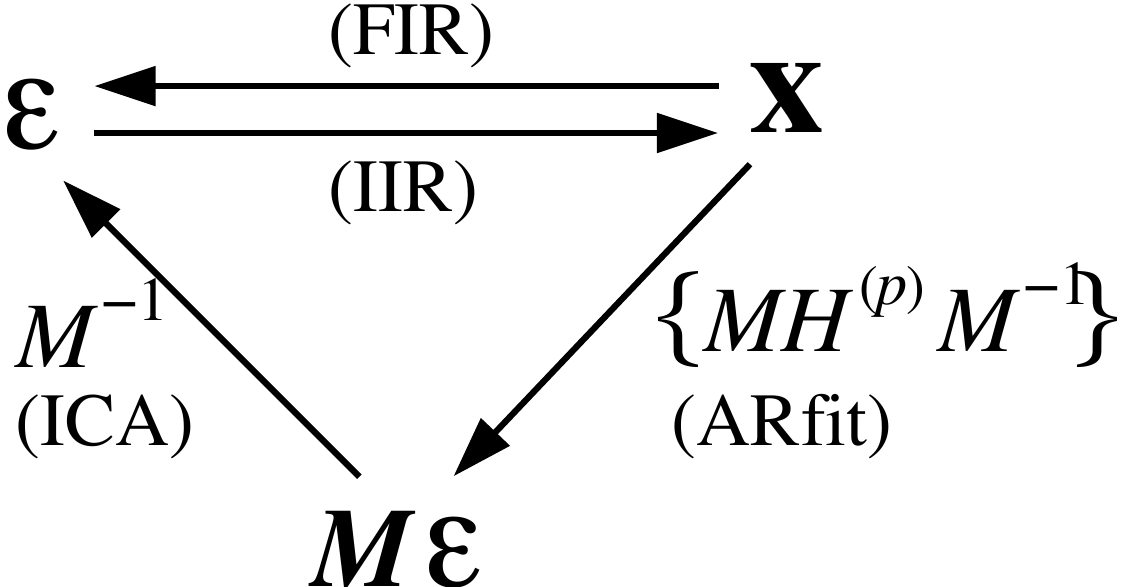}
\\
\vbox{\vskip5mm}(a) & & (b) \\
\vbox{\vskip18mm}
\includegraphics[width=0.17\textwidth]{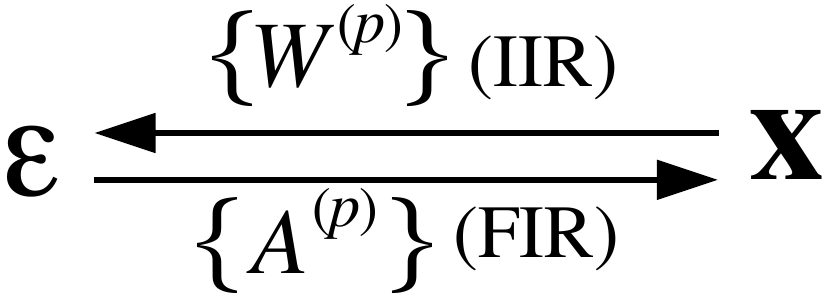}
& & \\
\vbox{\vskip5mm}(c) & &
\end{tabular}
\end{center}
\caption{Relations between (a) SCSA, (b) MVARICA and (c) CICAAR.
All approaches assume a non-Gaussian innovation sequence $\bf \varepsilon$.
SCSA and MVARICA fit an IIR model to the observed sequence $\bf x$,
while CICAAR assumes an FIR filter for it. Therefore, in SCSA and MVARICA
the inverse filter from $\bf x$ to the innovation $\bf 
\varepsilon$ is an FIR. MVARICA is a two step approach with AR fitting
to the observed sequence $\bf x$ and spartial demixing of the innovation
$M {\bf \varepsilon}$ obtained in the first step. On the other hand,
SCSA is a one-step approach which compute the inverse FIR filter by
convolutive ICA. We remark that the AR fitting in MVARICA relies only
on the second order statistics, which may cause the performance drops
compared to CSA.}
\label{fig:model_illu}
\end{figure}

It also turns out that the model Eq.~\eqref{eq:eegmodel} is very
similar to the convolutive ICA (cICA) \cite{attias98, parra00,
  anemuller03, cicaar} model. The only difference is that
Eq.~\eqref{eq:eegmodel} employs a FIR filter
to extract the innovations, while an infinite response filter (IIR) is
usually used in the cICA literature (see, e.g., \cite{cicaar}). This
discrepancy is explained by the different philosophies that are
associated with both methods. While in our approach the innovations
$\boldsymbol{\varepsilon}(t)$ arise as residuals of a finite-length
source-MVAR model, cICA understands them as sources of a finite-length
convolutional mixture. Nevertheless, our unregularized cost function
can be regarded as a maximum-Likelihood approach to an IIR version of
convolutive ICA. This leads us also to a new view of convolutive ICA
as performing an instantaneous demixing into correlated
sources. Hence, it is possible to conduct source connectivity analysis
using cICA (see Fig. \ref{fig:model_illu} for illustration).

Compared to MVARICA and time-domain implementations of convolutive ICA such as CICAAR \cite{cicaar}, our formulation has the advantage that sparse connectivity can easily be modeled by an additional penalty. This is not possible for CICAAR, because CICAAR only indirectly estimates the MVAR coefficients through their inverse filters. However, these are generally nonsparse, even if the true connectivity structure is sparse. Inverting the inverse coefficients is also generally not possible (recall, that convolutive ICA is equivalent to an infinite-length source-MVAR model). It is furthermore not possible to introduce a sparse regularization for MVARICA, since this method carries out the MVAR-estimation step in sensor-space, where no sparsity can be assumed.

By variation of the regularization parameter, our method is able to interpolate between a fully-correlated source model (comparable to convolutive ICA) and a model which allows no cross-talk between sources. Interestingly, the latter extreme can be seen as a variant of traditional instantaneous ICA, in which independence is measured in terms of mutual predictability with a Granger-type criterion. 

\subsection{Optimization}\label{sec:optimization}

\subsubsection{CSA}
The gradient of the unregularized cost function Eq.~\eqref{eq:negloglikelihood} is obtained as
%
\begin{eqnarray}
\lefteqn{\frac{\partial {\cal L}}{\partial W_d^{(p)}} \ = \ 
\delta(p)\left((P-T) {W^{(p)}}^{-\top} \vec{e}_d \right)} \nonumber \\
&& \hspace*{5mm} + \sum_{t = P+1}^T \tanh \left( \sum_{p=0}^P
	{W_{d}^{(p)}}^\top \vec{x}(t - p) \right) \vec{x}(t - p) \;, \quad
\end{eqnarray}
where $W_d^{(p)} := {W^{(p)}}^\top \vec{e}_d$, i.e. the $d$-th column vector
of ${W^{(p)}}^\top$.

We plug the gradient into a limited memory Broyden–-Fletcher–-Goldfarb–-Shanno (L-BFGS) optimizer \cite{lBFGS}\footnote{We use an implementation by Naoaki Okazaki, \texttt{http://www.chokkan.org/software/liblbfgs/}.} and observe that the algorithm always converges to the global optimum, while only the signs and order of the components may depend on the initialization. We use $W^{(0)} = I$ and $W^{(p)} = 0, p = 1, \hdots, P$ as a default initializer.

\subsubsection{SCSA via a modified L-BFGS algorithm}
Using sparse regularization, two additional difficulties emerge compared to the unregularized cost function. First, using the factorization Eq.~\eqref{eq:vartrans} the guaranteed convergence to the minimum observed for CSA is unlikely to be retained. Furthermore, the function Eq.~\eqref{eq:regnegloglikelihood} is not differentiable, when one of the terms $\| ( {H_{df}^{(1)}, \hdots, H_{df}^{(P)}} )^\top \|_2, d \neq f$ becomes zero, which is expected to be the case at the optimum.

For tackling these difficulties we here propose to use a modified version of the L-BFGS algorithm, which allows joint nonlinear optimization of $B$ and $\{H^{(p)} \}$, while taking special care of the nondifferentiability of the regularizer. The gradient of Eq.~\eqref{eq:regnegloglikelihood} is obtained as

\begin{eqnarray}\label{eq:gradientH}
\nonumber \frac{\partial {\cal L^{\mbox{\tiny SCSA}}}}{\partial
  H_{df}^{(p)}} & = & - \sum_{t = P+1}^T \tanh \left( 
{\rm s}_d (t) 
- \tilde{\rm s}_d(t) 
\right) 
\; {\rm s}_f(t-p)
\\
& & + \ \lambda \; \frac{H_{df}^{(p)}}{\left\| \left(H_{df}^{(1)}, \hdots, H_{df}^{(P)} \right)^\top \right\|_2}
\end{eqnarray}
and
\begin{eqnarray}\label{eq:gradientB}
\lefteqn{\frac{\partial {\cal L^{\mbox{\tiny SCSA}}}}{\partial B_d} \ = \ (P-T)
B^{-\top} \vec{e}_d} \nonumber \\
\nonumber & & \hspace*{8mm} 
+ \sum_{t = P+1}^T \sum_{d=1}^D \left\{ \tanh \left( 
{\rm s}_d(t) - \tilde{\rm s}_d(t)
\right) \vbox{\vskip8mm} \right. \nonumber \\
& & \left. \hspace*{15mm} \times \left( \vec{x}(t)
- \sum_{p = 1}^P {\rm x}_{d}(t-p) H_{d}^{(p)} \right) \right\} \quad \;.
\end{eqnarray}

Our modified L-BFGS algorithm checks before each gradient evaluation,
whether some terms $\| ({H_{df}^{(1)}}, \hdots, {H_{df}^{(P)}}
)^\top \|_2, d \neq f$ are already (close to) zero. If any of the terms
equals zero, the gradient is not defined uniquely but as a set
(subdifferential). Nevertheless it is straightforward to compute the
element of the subdifferential with minimum norm, whose sign inversion
is always a descent direction. Care must be taken because in practice we
would not find any of the above terms exactly equal to zero. Thus we
truncate the elements of $H$ corresponding to the terms with small norms
below some threshold to zero before computing the minimum norm
subgradient. If the minimum is indeed attained at the truncated point,
the minimum norm subgradient will be zero. Otherwise the 
subgradient will drive the solution out of zero. Further care must
be taken in practice to prevent the solution from oscillating in and out
of some zero.

We find that, using the outlined optimization procedure, sparse solutions can be found in shorter time, if the solution of the unregularized cost function is used as the initializer. The starting point can be obtained using the inverse transformation of Eq.~\eqref{eq:vartrans}, which is given by

\begin{eqnarray} \label{eq:invvartrans}
B & = & W^{(0)}\\
H^{(p)} & = & -W^{(p)}B^{-1}, \; p > 0 \;.
\end{eqnarray}

\subsubsection{SCSA via an EM algorithm}
Using joint optimization of $B$ and $\{H^{(p)} \}$, the heuristic pruning of connections might in some cases lead to suboptimal solutions regarding the composite cost function. For this reason, we present an alternative optimization scheme, which does not require any heuristic step. The idea here is to alternate between the estimation of both unknowns. Doing so can be justified as an application of the Expectation Maximization (EM) algorithm (see \cite{Neal98}).

Estimation of $B$ given $\{H^{(p)} \}$ (here called E-step) amounts to solving an unconstrained nonlinear optimization problem. Importantly, this problem is also convex, in contrast to the joint approach to SCSA parameter fitting. The convexity follows from the concavity of $\log |X|$ and $\log ( \sech (ax) )$ for constant $a$ (and from the fact that the sum of convex functions is convex.). The great advantage of convex problems is, that they feature a unique (local and global) minimum. In our case, the objective is smooth, so the minimum is guaranteed to be found by the L-BFGS algorithm, making use of the gradient in Eq.~\eqref{eq:gradientB}.

Optimization with respect to $\{H^{(p)} \}$ for fixed $B$ (M-step) is
more involved, since the nondifferentiable Group Lasso regularizer
remains. Smooth optimization methods like L-BFGS are unlikely to find the
exact solution here. However, this problem is not as difficult as the
joint optimization problem, since it is convex. This can be seen from
the fact that it is composed of a sum of $- \log ( \sech (ax) )$ terms
(loss function) and the Group Lasso term (regularizer), which is a sum
of $\ell_2$-norms and thus convex. Hence we can solve this problem using
the Dual Augmented Lagrangian (DAL) procedure \cite{dal09}, which has
recently been introduced as a method for minimizing arbitrary convex
loss functions with additional Lasso or Group Lasso
penalties. Application of DAL requires the loss function and its
gradient, the convex conjugate (Legendre transform) of the loss function, as well as gradient and Hessian of the conjugate loss. Let $\vec{s}(t) = B \vec{x}(t)$ be the demixed sources and $\tilde{\vec{s}}(t) = \sum_{p=1}^P H^{(p)} \vec{s}({t-p}) $ be their autoregressive approximations. The loss function in terms of $\tilde{\vec{s}}$ is defined as 

\begin{eqnarray}\label{eq:mloss}
{\cal L}^{\mbox{\tiny M}}(\tilde{\vec{s}}) & = & - \sum_{t = P+1}^T
\sum_{d=1}^D \log \left( \frac{1}{\pi} \sech \left(  \tilde{\rm
  s}_d(t)-{\rm s}_d(t) \right) \right)\,.\qquad
\end{eqnarray}
The gradient is 
\begin{eqnarray}\label{eq:mlossgrad}
\frac{\partial {\cal L^{\mbox{\tiny M}}}}{\partial \tilde{\rm s}_d(t)}
& = & \tanh(\tilde{\rm s}_d(t)-{\rm s}_d(t)) \;.
\end{eqnarray}

Let ${\rm a}_d(t)$ ($d=1,\ldots,D$, $t=P+1,\ldots,T$) denote the dual
variables associated with the Legendre transform. The conjugate loss
function is defined on the interval $[-1, 1]$ and evaluates to 
\begin{align}\label{eq:mlossdual}
\lefteqn{{\cal D}^{\mbox{\tiny M}}(\vec{a})} \nonumber \\
 & = \sum_{t = P+1}^T
 \sum_{d=1}^D
\sup_{\tilde{\rm s}_d(t)}\left({\rm a}_d(t)\tilde{\rm s}_d(t)-\log\frac{\sech\left(
\tilde{\rm s}_d(t)-{\rm s}_d(t)
\right)}{\pi}\right) \nonumber \\
&=\sum_{t = P+1}^T \sum_{d=1}^D\Biggl(
 \frac{1-{\rm a}_d(t)}{2}\log\frac{1-{\rm a}_d(t)}{2} \nonumber \\
&\qquad+\frac{1+{\rm a}_d(t)}{2}\log\frac{1+{\rm a}_d(t)}{2}
- {\rm a}_d(t) {\rm s}_d(t) +  \log \frac{2}{\pi}\Biggr)\;.
\end{align}

The gradient of the conjugate loss is given by
\begin{eqnarray}\label{eq:mlossdualgrad}
\frac{\partial {\cal D^{\mbox{\tiny M}}}(\vec{a})}{\partial {\rm
    a}_d(t)} & = & \frac{1}{2}\log\frac{1+{\rm a}_d(t)}{1-{\rm
    a}_d(t)}- {\rm s}_d(t) \;.
\end{eqnarray}
The Hessian is diagonal with elements
\begin{eqnarray}\label{eq:mlossdualhess}
\frac{\partial^2 {\cal D^{\mbox{\tiny M}}}(\vec{a})}{\partial {\rm
    a}_d(t)^2} & = & \frac{1}{2(1 - {\rm a}^2_d(t))} \;.
\end{eqnarray}

Having defined the E- and M-steps, we have turned a nonconvex estimation problem into a sequence of two convex problems, which can both be solved exactly. A final estimate of the model parameters can now be obtained by alternating between E- and M-steps until convergence.

\subsection{Treating source autocorrelations}
Diagonal parts of the MVAR matrices $\{H^{(p)} \}$ model the sources' autocorrelation and should preferably not be pruned. However, in some cases numerical stability can be increased if these variables are also penalized, especially if $D$ and $P$ are large. For this reason, we use a slight variation of the cost function Eq.~\eqref{eq:regnegloglikelihood} in practice, which includes
\begin{equation}
\left\| \left( {H_{11}^{(1)}}, \hdots, {H_{11}^{(P)}}, \hdots, {H_{DD}^{(1)}}, \hdots, {H_{DD}^{(P)}} \right)^\top \right\|_2\\
\end{equation}
as an additional penalty term. 
 The augmented objective function can be minimized using the techniques presented in Section \ref{sec:optimization}.

\section{Performance under realistic conditions}
\label{sec:experiment}
We conducted the following simulations in order to assess the
performance of the proposed source connectivity analysis compared to
those of existing approaches. 

\subsection{Data generation}
We simulated seven time-series (pseudo-sources) of length $N = 2000$ according to an MVAR model of order $P = 4$. Seven out of the forty-two possible interactions were modeled by allowing the corresponding offdiagonal MVAR coefficients $H_{df}^{(p)}, d \neq f, 1 \leq p \leq P$ to be nonzero. The innovations were drawn from the $\sech$-distribution (Note that the assumption of non-Gaussianity is crucial for recovering mixed sources.). 

The pseudo-sources were mapped to 118 EEG channels using the theoretical spread of seven randomly placed dipoles. The spread was computed using a realistic forward model \cite{nolteleadfield} which was built based on anatomical MR images of the ``Montreal head'' \cite{montreal}. See Fig.~\ref{fig:toy_reconstruction} for an example illustrating the data generation.

In reality, measurements are never noise-free and the following model holds rather than Eq.~\eqref{eq:eegmodel}

\begin{eqnarray}\label{eq:noisyeegmodel}
\vec{x}(t)& = & M \vec{s}(t) + \boldsymbol{\xi}(t) \;.
\end{eqnarray}

Since none of the methods compared here (see below) explicitly models a noise term, it is important to evaluate their robustness to model violation. To this end, we constructed additional variants of the pseudo-EEG dataset by adding six different types of noise $\boldsymbol{\xi}$. The six variants (N1-N6) are summarized in TABLE~\ref{tab:noise}. These variants differ in their degree of spatial and temporal correlation as follows. In variants N1 and N4, $\xi_i(t), i = 1, \hdots, M$ were drawn independently for each sensor, i.e., have no spatial correlation. For variants N2 and N5 noise terms $\xi_i^*(t), i = 1, \hdots, M$ were drawn independently for each \emph{source}. In this case, sources and noise contributions to the EEG share the same covariance given by the mixing matrix $M$, i.e., $\vec{x}(t) = M (\left( \vec{s}(t) + \boldsymbol{\xi}^*(t) \right) $. For the last variants N3 and N6, spatially independent noise sources were simulated at all nodes of a grid covering the whole brain, yielding the model $\vec{x}(t) = M \vec{s}(t) + M^* \boldsymbol{\xi}^*(t)$. Here, in contrast to the previous model, noise contributions are not collinear to the sources. We further distinguish between noise sources with and without temporal structure. In variants (N1-N3), noise terms were drawn $i.i.d.$ from a normal distribution at each time instant $t$. In variants N4-N6, the temporal structure was determined by a univariate AR model of order 20, i.e., $\boldsymbol{\xi}^*(t) = \sum_{p=1} ^{20} {H^*}^{(p)} \boldsymbol{\xi}^*({t-p}) + \boldsymbol{\varepsilon}^*(t)$.

Note that, since no time-delayed dependencies between noise sources were modeled, no additional Granger-causal effects were introduced by the noise. We used a signal-to-noise ratio (SNR) of 2 in all experiments, where SNR is defined as 

\begin{equation}
\mbox{SNR} = \frac{\left\| M \left( \vec{s}(1), \hdots, \vec{s}(T) \right) \right\|_{\cal F}}{\left\| \left( \boldsymbol{\xi}^*(1), \hdots, \boldsymbol{\xi}^*(T) \right) \right\|_{\cal F}} \;,
\end{equation}
and $\| \cdot \|_{\cal F}$ is the Frobenius norm of a matrix.

Finally, PCA was applied to the pseudo-EEG to reduce the dimensionality to $D = 7$ (the original number of sources) by taking just the seven strongest PCA components. One-hundred datasets with different realisations of MVAR coefficients, innovations and noise were constructed for each category.

\begin{table}[htb]
\caption{\label{tab:noise} The six types of noise used in the simulations. Noise with
  temporal correlation structure was created using univariate AR
  models of order 20. Spatial correlation was introduced using the
  forward model. We distinguish between the case, where noise sources
  coincide with the true dipoles (\textsuperscript{a}) and the case in
  which noise from all brain sites contributes to the measurements
  (\textsuperscript{b})}.
\begin{center}
\begin{tabular}{lcc}
\toprule
& independent in time & correlated  in time \\
\midrule
independent in sensors & N1 & N4 \\
correlated in sensors\textsuperscript{a} & N2 & N5 \\
correlated in sensors\textsuperscript{b} & N3 & N6 \\
\bottomrule
\end{tabular}
\end{center}
\end{table}

\subsection{Methods}
We tested the ability of ICA, MVARICA, CICAAR and the two proposed methods CSA and SCSA to reconstruct the seven sources and their connectivity structure. Although the goal of instantaneous ICA is fundamentally different to source connectivity analysis, it was also included here in the comparison. This is since, even if independence of the sources is not fulfilled, ICA might still provide as-least-as-possible dependent components, the connectivity of which might be analyzed. The ICA variant used here is based on temporal decorrelation \cite{molgedey94, belouchrani, Ziehe98, ZieMueNolMacCur00} (implemented by fast approximate joint diagonalization \cite{ffdiag}). The number of temporal lags was set to 100.

MVARICA, CICAAR, CSA and SCSA were tested with $P \in \{ 1, 2 ,
\hdots, 7\}$ temporal lags, where four is the true MVAR model order
for CSA, SCSA and MVARICA. CICAAR has the disadvantage here, that it
may generally require extended temporal filters for reconstructing
sources following model Eq.~\eqref{eq:eegmodel}. However, due to
computation time constraints, $P=7$ was taken as the maximum lag also
for this method. For MVARICA and CICAAR, we used implementations
provided by the respective authors. These implementations adopt the
Bayesian Information Criterion (BIC) for selecting the appropriate
number of time lags. The same criterion was used to select the model
order in CSA and SCSA. The regularization constant $\lambda$ of SCSA
was set by $5$-fold cross-validation. SCSA estimates of $\{H^{(p)} \}$ and $B$ were obtained either jointly using the modified L-BFGS algorithm or alternately using 20 additional EM steps. These variants are named SCSA and SCS\_EM here, respectively.

\subsection{Performance measures}
The most important performance criterion is the reconstruction of the mixing matrix, since all other relevant quantities can basically be derived from it. All considered methods provide an estimate $\hat M^{-1}$ of the demixing, which can be inverted to yield an estimated mixing matrix. The columns of the mixing matrix correspond to spatial field patterns of the estimated sources, but unfortunately these patterns can generally only be determined up to sign, scale and order. For this reason, optimal pairing of true and estimated patterns as described in \cite{pairing04} was performed. The similarity measure between patterns was slightly modified compared to the one used in \cite{pairing04}. We used the goodness-of-fit achieved by a linear least-squares regression of one to another pattern. For a true pattern $M_{d}$ and an estimated pattern $\hat M_{f}$ the optimal regression coefficient is 

\begin{equation}
  c\left(M_{d}, \hat M_{f}\right) = \frac{\hat M_{f}^\top M_{d}}{\| \hat M_{f} \|^2} \\
\end{equation}

and the goodness-of-fit (GOF) is 

\begin{equation}
  \mbox{GOF}\left(M_{d}, \hat M_{f}\right) = \frac{\|c \hat M_{f} - M_{d}\|}{\| M_{d} \|} \;.
\end{equation}

\begin{figure}[htb]
\begin{center}
\subfigure[Simulated Dipole]{\includegraphics[width=0.2\textwidth]{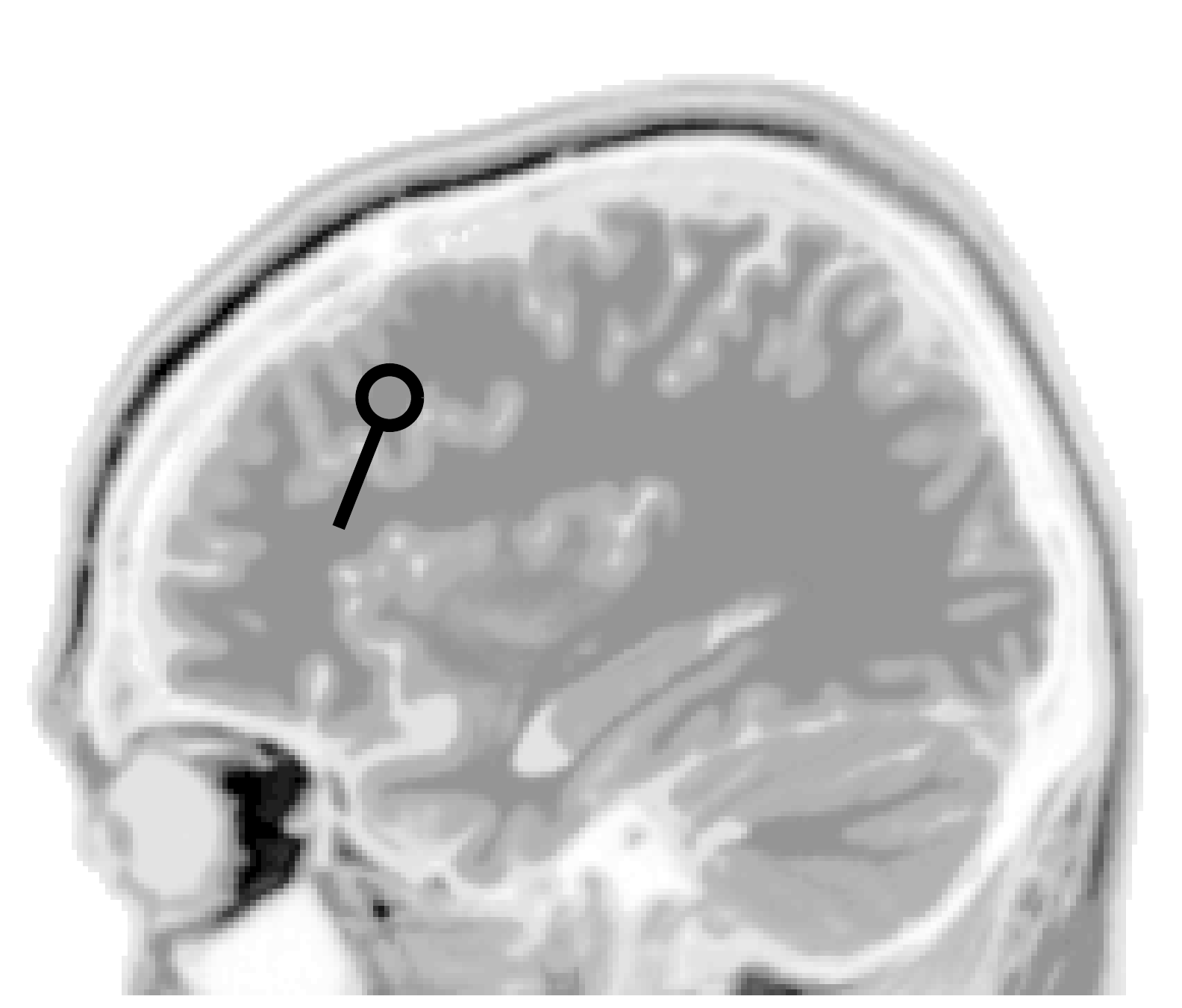}}
\subfigure[Corresponding EEG Pattern]{\includegraphics[width=0.2\textwidth]{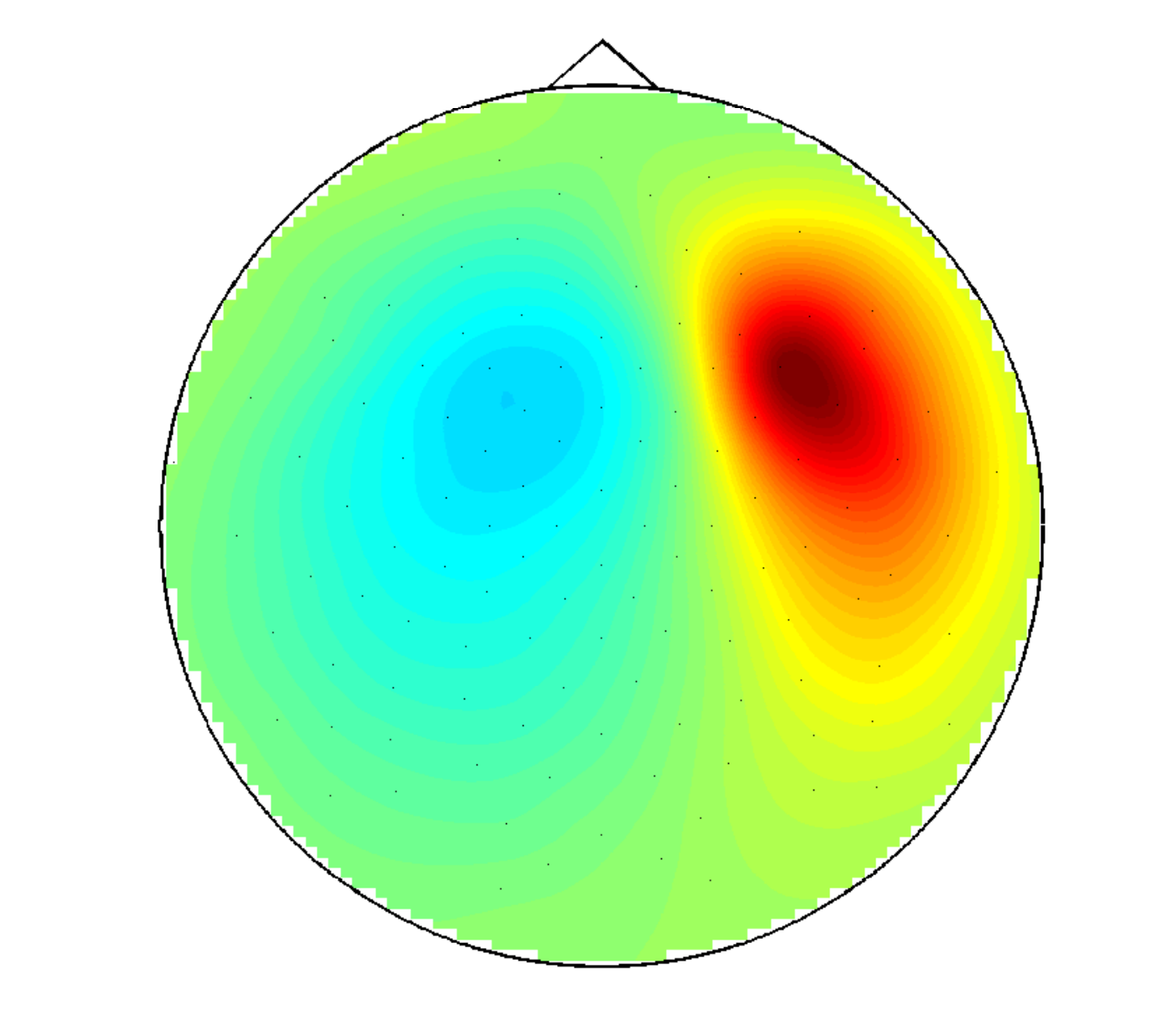}}\\
\subfigure[Estimated Dipole]{\includegraphics[width=0.2\textwidth]{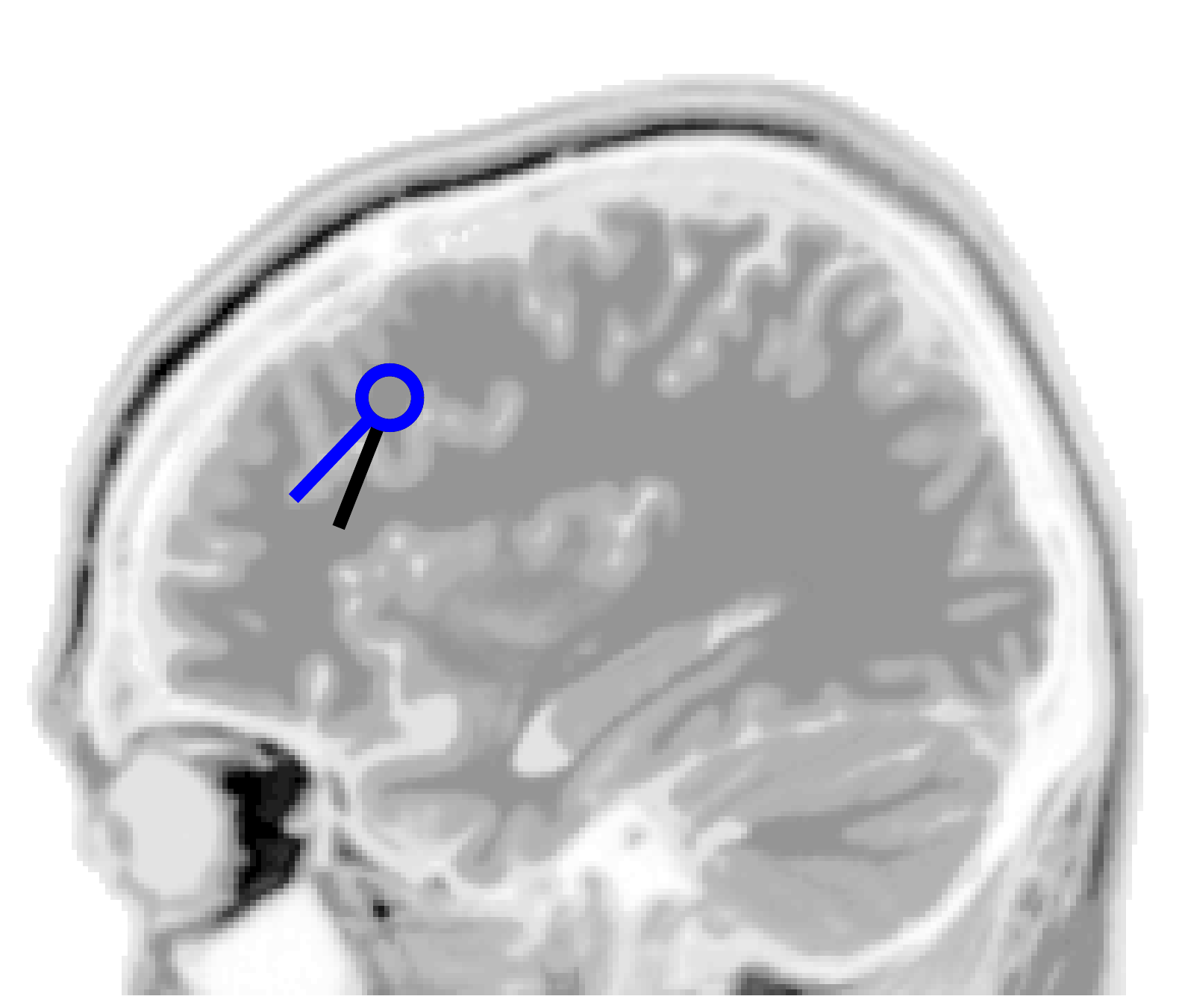}}
\subfigure[Estimated EEG Pattern]{\includegraphics[width=0.2\textwidth]{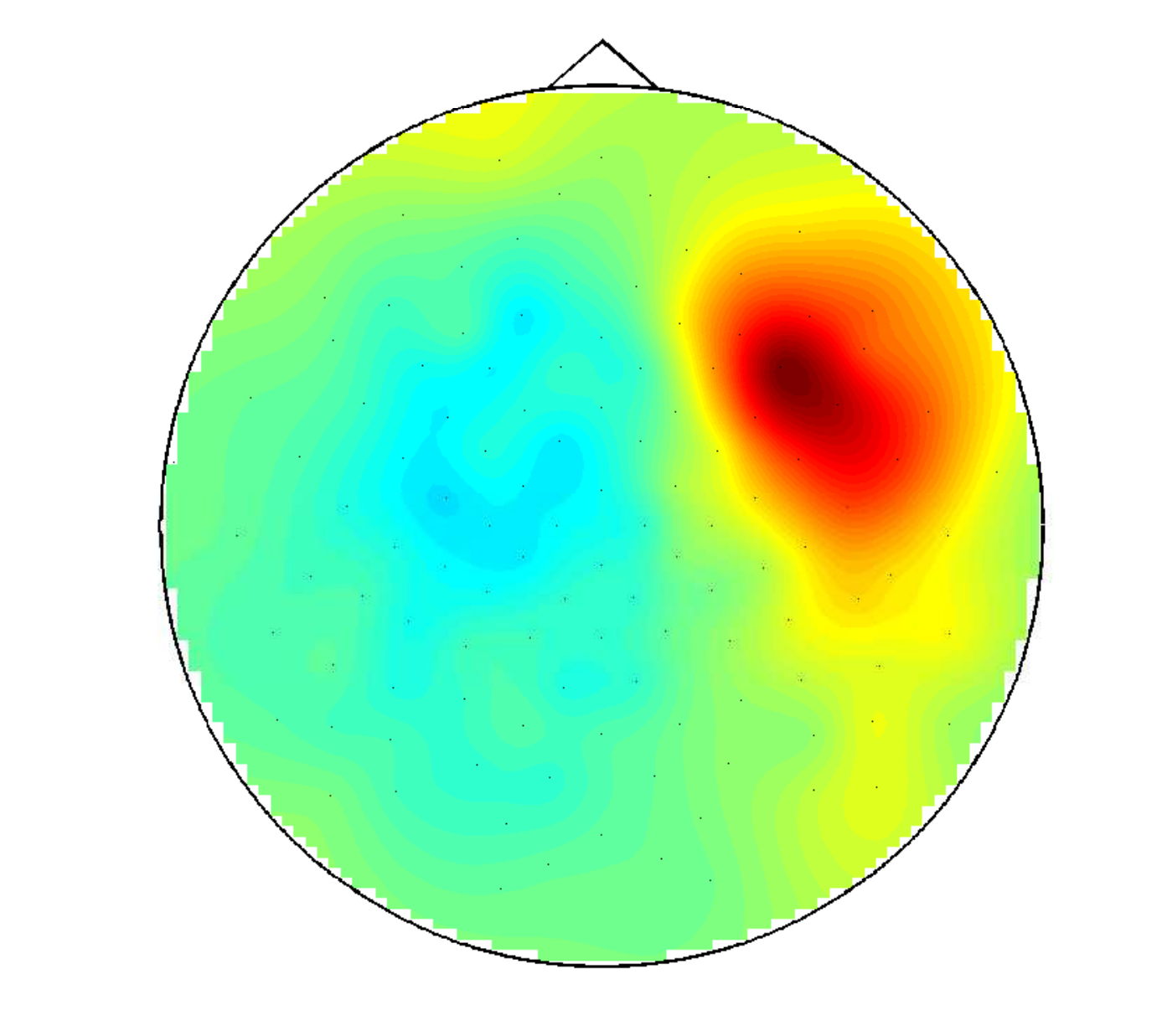}}
\end{center}
\caption{Example of simulated data (noise type N1) and corresponding reconstruction by SCSA. (a) Simulated dipole. (b) Field pattern describing the dipole's influence on the EEG (one column of $M$). (d) Field pattern as estimated by SCSA from noisy EEG time series. (c) Reconstructed dipole, obtained from the estimated pattern.}
\label{fig:toy_reconstruction}
\end{figure}

Having found the optimal pairing, the colums of $M$ were permuted and scaled to approximate $M$ as good as possible using the optimal regression coefficients. The goodness-of-fit with respect to the whole matrix $M$ was used to evaluate the quality of the different decompositions. Additionally, using the optimally-matched mixing patterns, dipole scans were conducted and the deviation of the obtained dipole locations from the true ones was measured. A typical example of a mixing pattern estimated by SCSA and the corresponding reconstructed dipole is shown in Fig.~\ref{fig:toy_reconstruction}. 

Finally, causal discovery according to \cite{HauNolMueKra08} was carried out on the demixed sources. The exact technique used was MVAR estimation with Ridge Regression. For the MVAR parameters estimated by Ridge Regression an approximate multivariate Gaussian distribution can be derived, which was used to test the coefficients for beeing significantly different from zero. An influence from $s_i$ to $s_j$ was defined, if the $p$-value of one of the coefficients $H_{ij}^{(p)}, p = 1, \hdots, P$ fell below the critical value. As a third performance criterion, the area under curve (AUC) score for correctly discovering the interaction structure was calculated by varying the significance threshold and comparing estimated and true connectivity matrix for each threshold. Note that this way of connectivity estimation was pursued here in order to treat all methods equally. This was necessary, since not all methods provide built-in connectivity estimates. For SCSA, however, interaction analysis could as well have been done by directly examining MVAR coefficients. Note further, that using Ridge Regression based testing, the non-Gaussianity of the source MVAR innovations is only indirectly used through the use of the demixing matrix, but not for actual MVAR estimation. For this reason, the MVAR coefficients directly estimated by SCSA may be preferred to a subsequent Ridge Regression step when using SCSA in practice.

\subsection{Results}
\begin{figure}[htb]
\begin{center}
\includegraphics[width=0.45\textwidth]{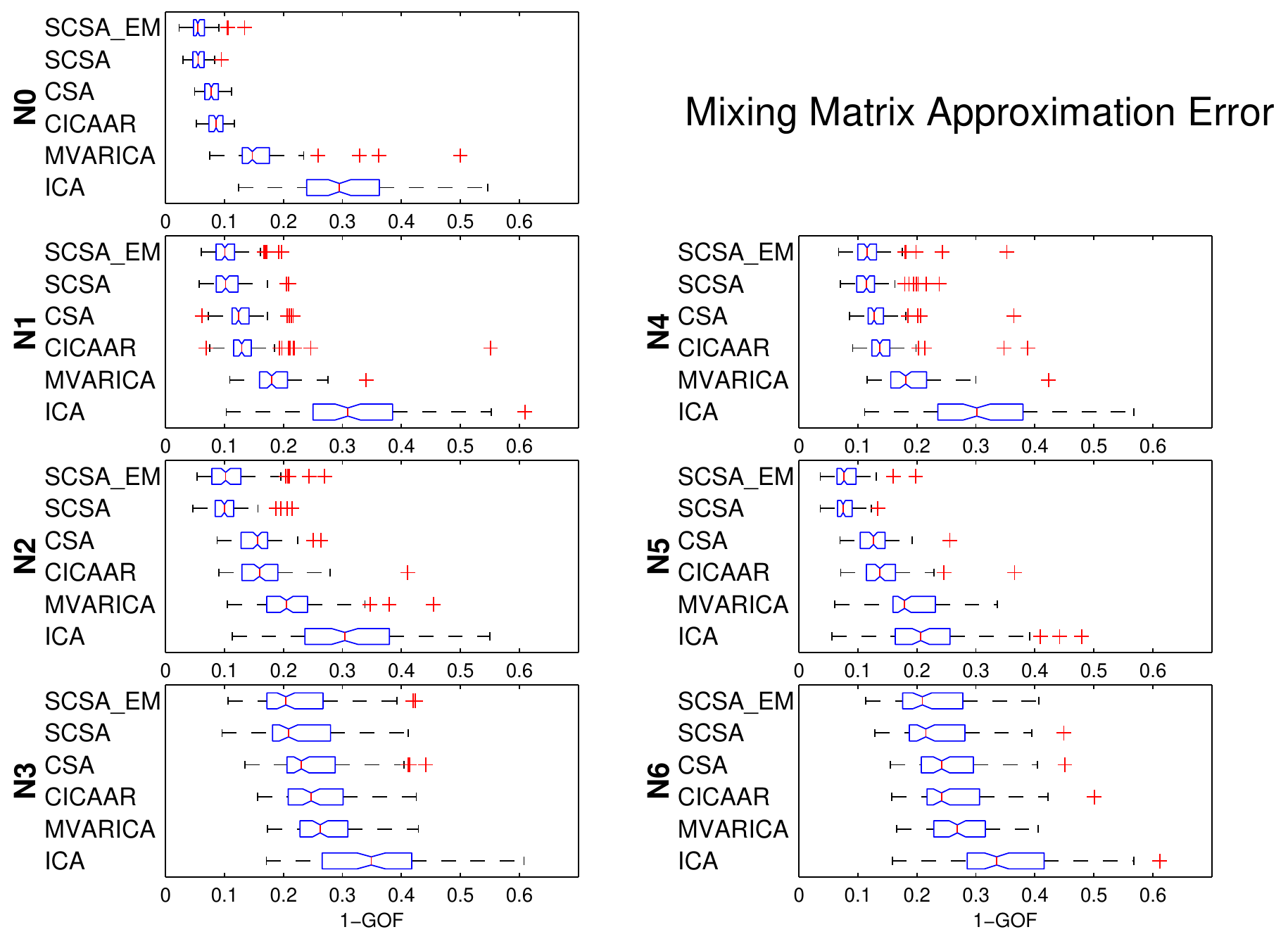}
\end{center}
\caption{Estimation errors of the mixing matrix according to the goodness-of-fit (GOF) criterion. Results are shown for the proposed (Sparsely-) Connected Sources Analysis variants (SCSA\_EM, SCSA, CSA) and three alternative approaches (CICAAR, MVARICA, ICA). Different subfigures depict the methods' performance in the noiseless cass (N0), as well as in the presence of different types of noise (N1-N6, see TABLE~\ref{tab:noise}).}
\label{fig:mix_error}
\end{figure}

Fig. \ref{fig:mix_error} shows, how well the mixing matrix was approximated by the different approaches. One boxplot is drawn for the noiseless case (N0) and each of the six noisy variants (N1-N6, see Table \ref{tab:noise}). The plots show the median performance over 100 repetitions, as well as the lower and upper quartiles and the extremal values. Outliers (red crosses) were removed. As a result of the simulations, SCSA typically achieves the smallest reconstruction error, followed by CSA, CICAAR, MVARICA and ICA. In many cases, these differences are also significant, as indicated by notches in the boxes.

\begin{figure}[htb]
\begin{center}
\includegraphics[width=0.45\textwidth]{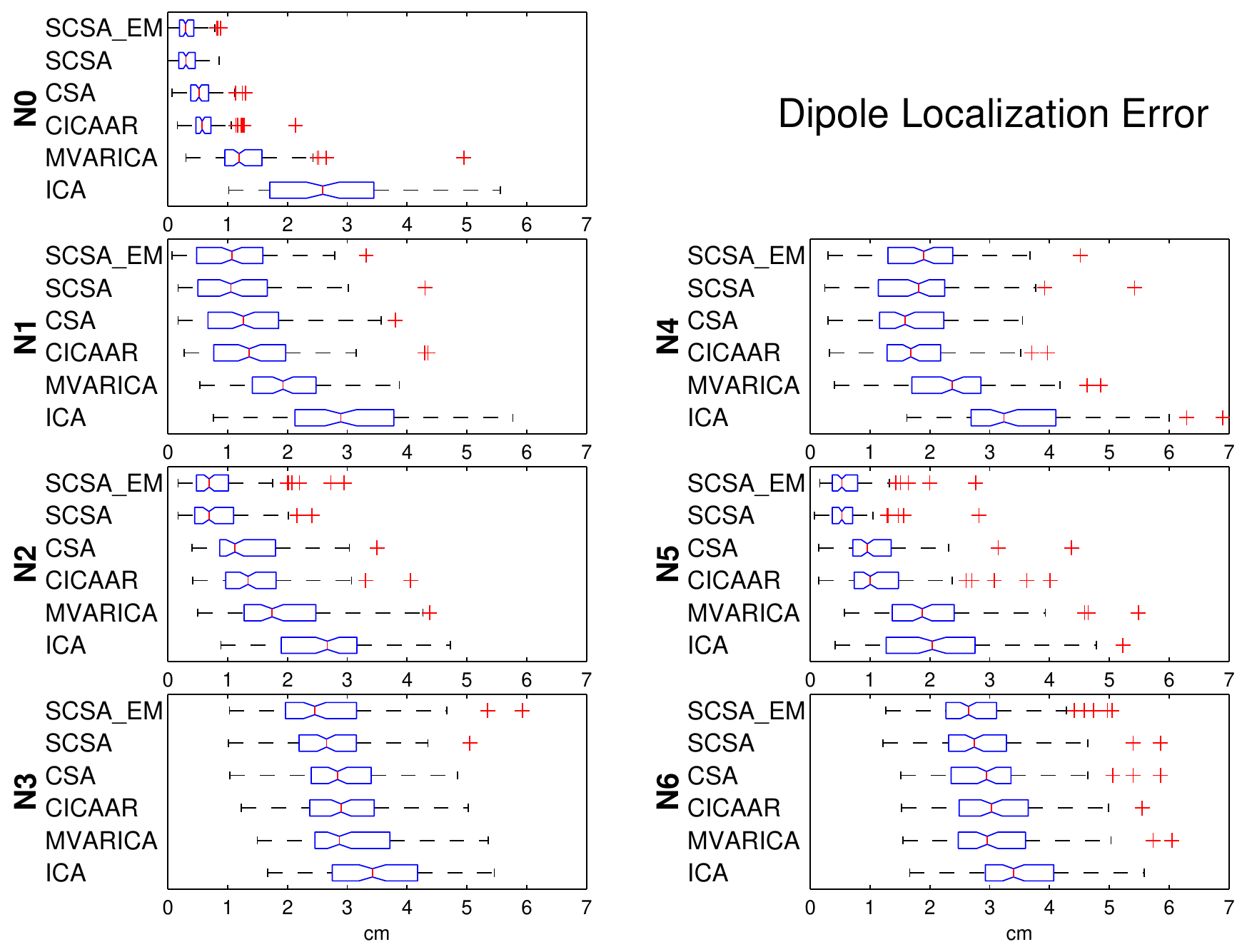}
\end{center}
\caption{Localization errors of dipole fits conducted on the estimated mixing field patterns. Results are shown for the proposed (Sparsely-) Connected Sources Analysis (SCSA\_EM, SCSA, CSA) variants and three alternative approaches (CICAAR, MVARICA, ICA). Different subfigures depict the methods' performance in the noiseless cass (N0), as well as in the presence of different types of noise (N1-N6, see TABLE~\ref{tab:noise}).}
\label{fig:dip_error}
\end{figure}

\begin{figure}[htb]
\begin{center}
\includegraphics[width=0.45\textwidth]{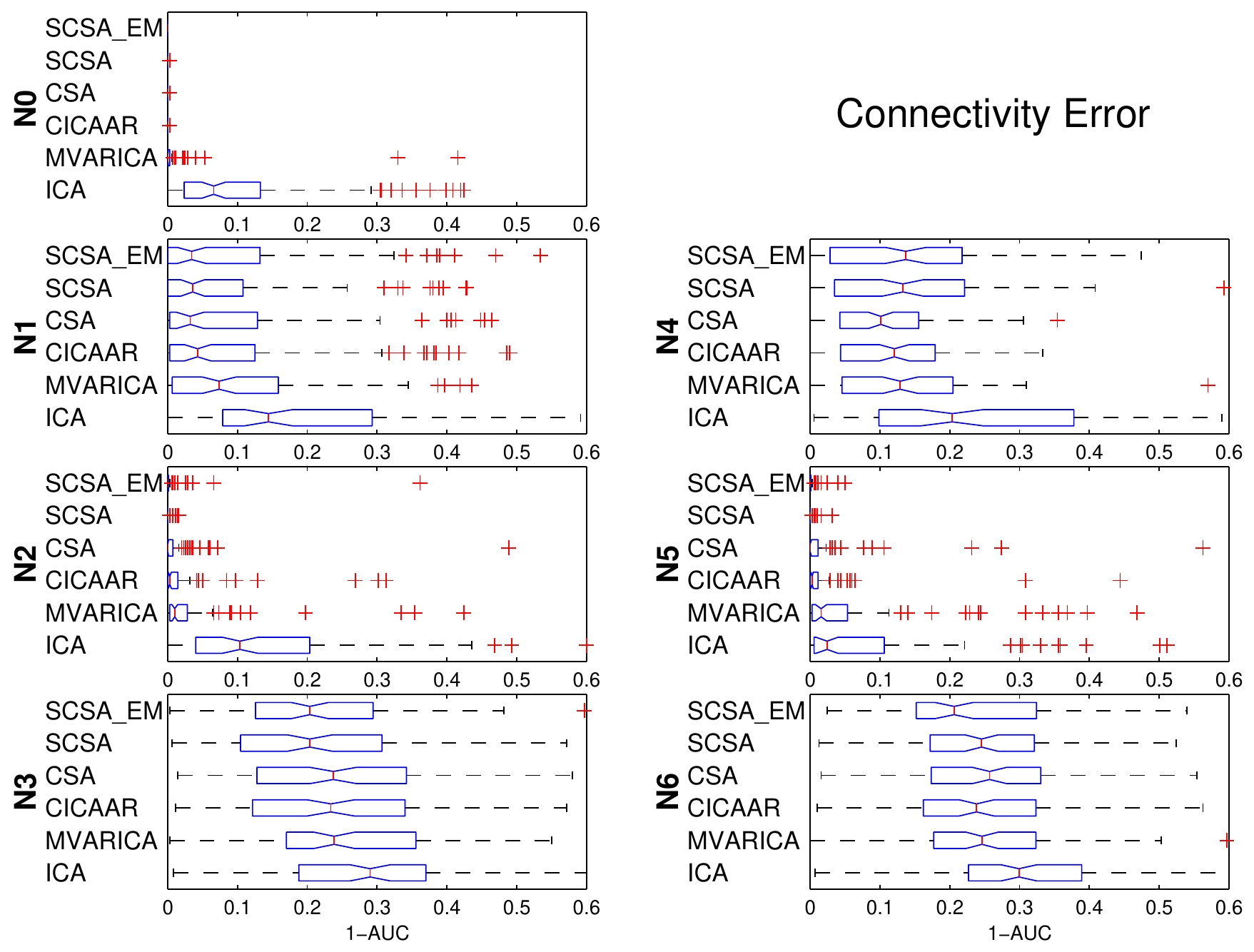}
\end{center}
\caption{Estimation errors regarding the source connectivity structure as measured by fitting an MVAR model subsequently to the demixed sources and testing the obtained coefficients for significant interaction. The performance measure reported is the area under the curve (AUC) score obtained by varying the significance level.}
\label{fig:conn_error}
\end{figure}

Correct (de-)mixing matrix estimation affects both the localization error achievable by applying inverse methods to the estimated patterns and the error of any connectivity analysis performed at the demixed sources. As a result of good mixing matrix approximation, SCSA also achieves smaller dipole localization errors than all other methods, except in one scenario (shown in Fig. \ref{fig:dip_error}). The same situation occurs when it comes to estimating the connectivity between sources (Fig.~\ref{fig:conn_error}). 

Interestingly, the higher numerical stability we observed for the EM variant of SCSA compared to joint parameter estimation only sometimes leads to superior performance. This may be related to our observation, that the difference between the implementations becomes large only for excessively large amounts of regularization, which are not optimal in terms of the cross-validation criterion. Another reason might be that the instability of the MVAR coefficients around zero does not play a crucial role in our current evaluation, since all performance measures used here were solely derived from the demixing matrix.

Regarding noise influence it might be said that the relative degradation of performance in the presence of noise is the same for all methods. Generally, noise that is collinear to the sources (N2/N5) seems to be less problematic than noise that is uncorrelated across sensors (N1/N4) and noise with arbitrary spatial correlation structure (N3/N6). Judging from mixing matrix approximation and dipole localization errors, the temporal structure of the noise seems not to affect the performance much. However, small errors in the (de-)mixing matrix can have quite a negative effect on the connectivity estimation, as can be seen in the right part of Fig.~\ref{fig:conn_error}.

\begin{figure}[htb]
\begin{center}
\includegraphics[width=0.45\textwidth]{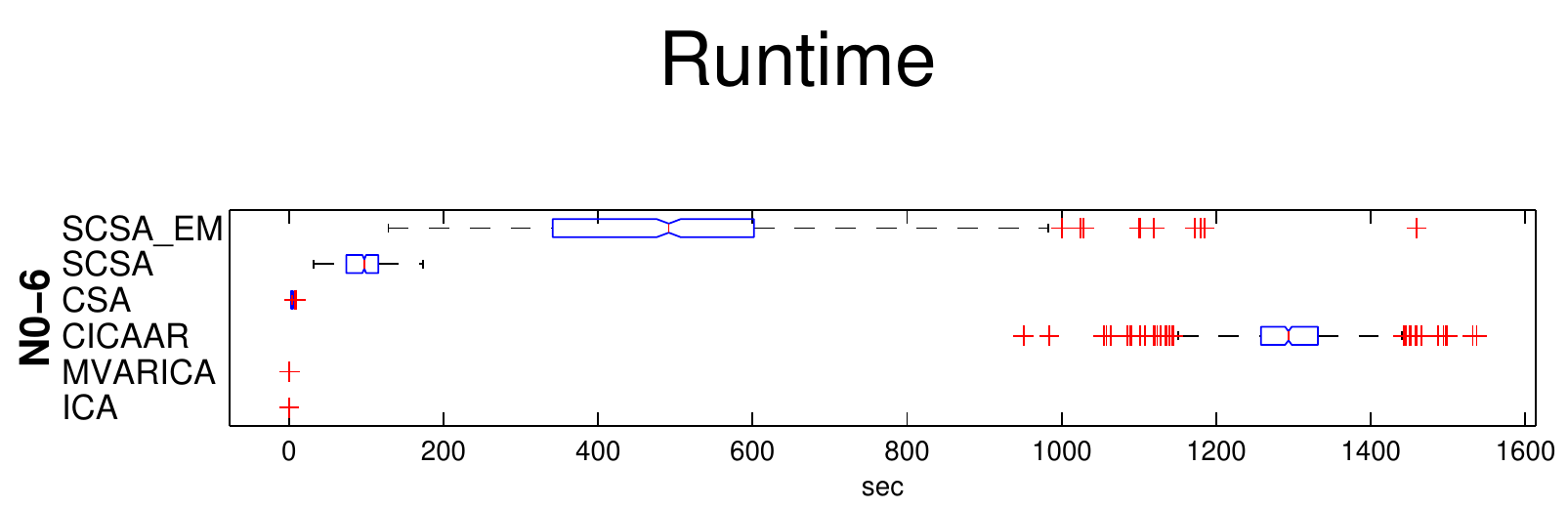}
\end{center}
\caption{Average runtime of the proposed (Sparsely-) Connected Sources Analysis variants (SCSA\_EM, SCSA, CSA) and three alternative approaches (CICAAR, MVARICA, ICA), taken over all experiments conducted for this study.}
\label{fig:runtime}
\end{figure}

The time each method consumed on average for processing one dataset is shown in Fig.~\ref{fig:runtime}. Most methods finish in rather short time, while the EM implementation of SCSA is in medium range and CICAAR requires the longest time. However, for SCSA there is still room for improvement, since the regularization parameter of this method is currently selected by the cross-validation procedure, which could be changed. 

\section{Discussion}\label{sec:discussion}

Let us recall the assumptions we make to identify
individual brain sources and to estimate their
interactions. While ICA results
in a unique decomposition assuming statistical independence, 
such an assumption is inconsistent when studying brain interactions.
However, all neural interactions require a minimum delay well within
the temporal resolution of electrophysical measurements of brain activity.
Hence, it makes sense to assume independent innovation processes
and to model all interactions explicitly using AR matrices.
In relation to ICA we pay some price for that: In our case, independence
is exploited effectively on reduced information contained in the
residuals of the model. In principle, this can be a cause for less
stable estimates. To increase stability, we have included sparsity
assumptions based on the idea that only a few brain connections
can be as strong to be observable in EEG data which is especially
the case in the presence of artifacts and background noise.

We emphasize that we assume a linear dynamical model and non-Gaussian
innovation processes, i.e. the only cause of non-Gaussianity is
the innovation process itself. Real brain networks are, of course, more
complicated. However, the question
whether nonlinear dynamical models may improve the results
or are even essential for a correct decomposition is beyond the scope
of this paper and will be addressed in the future.
Similarly, we assumed the total number of sources to be less or 
equal the number of channels. Apparently, the significance of this 
problem decreases when using a large number of channels. 

\section{Conclusion}
\label{sec:conclusion}
Analysing the functional brain connectivity is a hard problem, since
volume conduction effects in EEG/MEG measurements can give rise to
spurious conductivity. In this work we have established a novel
connectivity analysis method SCSA that overcome these problems in an
elegant and numerically appealing manner using Group Lasso. In detail,
EEG is modeled as a linear mixture of correlated sources, then we
estimate jointly the demixing process and the MVAR model (which is the
model basis for the correlated sources). For this we assume that the
innovations driving the source MVAR process are super-Gaussian
distributed and (spatially and temporally) independent. To avoid
overfitting we regularize the model using the Group Lasso penalty. In
this manner we can achieve a data driven interpolation between two
extremes: a source model that has full correlations, i.e.~convolutive
ICA and conventional ICA that does not allow for cross-talk between the
extracted sources. In between, our method extracts a sparse
connectivity model. We demonstrate the usefulness of SCSA with
simulated data, and compare to a number of existing algorithms with
excellent results.

Future work will study the link between methods for compensating
non-stationarity in data such as Stationary Subspace Analysis
(SSA, \cite{BunMeiKirMue09}) and our novel connectivity assessment. In addition,
we aim to localize the extracted components of connectivity using distributed source models to enhance
physiological interpretability (e.g.~\cite{neuroimage08,nips08}).

\section*{Acknowledgement}
This work was partly supported by the \emph{Bundesministerium f\"ur
Bildung und Forschung} (BMBF), Fkz~01GQ0850 and by the European ICT
Programme Project FP7-224631 and 216886. We thank Germ\'an G\'omez Herrero and Mads Dyrholm for making the source code of their algorithms available, and Nicole Kr\"amer for discussions.

\bibliographystyle{IEEEtran}
\bibliography{lit}

\begin{thebibliography}{10}
\providecommand{\url}[1]{#1}
\csname url@samestyle\endcsname
\providecommand{\newblock}{\relax}
\providecommand{\bibinfo}[2]{#2}
\providecommand{\BIBentrySTDinterwordspacing}{\spaceskip=0pt\relax}
\providecommand{\BIBentryALTinterwordstretchfactor}{4}
\providecommand{\BIBentryALTinterwordspacing}{\spaceskip=\fontdimen2\font plus
\BIBentryALTinterwordstretchfactor\fontdimen3\font minus
  \fontdimen4\font\relax}
\providecommand{\BIBforeignlanguage}[2]{{%
\expandafter\ifx\csname l@#1\endcsname\relax
\typeout{** WARNING: IEEEtran.bst: No hyphenation pattern has been}%
\typeout{** loaded for the language `#1'. Using the pattern for}%
\typeout{** the default language instead.}%
\else
\language=\csname l@#1\endcsname
\fi
#2}}
\providecommand{\BIBdecl}{\relax}
\BIBdecl

\bibitem{NolZieNik08}
G.~Nolte, A.~Ziehe, V.~V. Nikulin, A.~Schl\"ogl, N.~Kr\"amer, T.~Brismar, and
  K.~R. M\"uller, ``{{R}obustly estimating the flow direction of information in
  complex physical systems},'' \emph{Phys. Rev. Lett.}, vol. 100, p. 234101,
  Jun 2008.

\bibitem{granger}
C.~Granger, ``Investigating causal relations by econometric models and
  cross-spectral methods,'' \emph{Econometrica}, vol.~37, pp. 424--438, 1969.

\bibitem{dtf}
M.~J. Kaminski and K.~J. Blinowska, ``{{A} new method of the description of the
  information flow in the brain structures},'' \emph{Biol Cybern}, vol.~65, pp.
  203--210, 1991.

\bibitem{pdc}
L.~A. Baccal\'a and K.~Sameshima, ``{{P}artial directed coherence: a new
  concept in neural structure determination},'' \emph{Biol Cybern}, vol.~84,
  pp. 463--474, Jun 2001.

\bibitem{HauNolMueKra08}
S.~Haufe, G.~Nolte, K.-R. M\"uller, and N.~Kr\"amer, ``Sparse causal discovery
  in multivariate time series,'' in \emph{Proceedings of the NIPS'08 Causality
  Workshop}, 2009.

\bibitem{imagcoh}
G.~Nolte, O.~Bai, L.~Wheaton, Z.~Mari, S.~Vorbach, and M.~Hallett,
  ``{{I}dentifying true brain interaction from {E}{E}{G} data using the
  imaginary part of coherency},'' \emph{Clin Neurophysiol}, vol. 115, pp.
  2292--2307, Oct 2004.

\bibitem{pmid17894381}
A.~G. Guggisberg, S.~M. Honma, A.~M. Findlay, S.~S. Dalal, H.~E. Kirsch, M.~S.
  Berger, and S.~S. Nagarajan, ``{{M}apping functional connectivity in patients
  with brain lesions},'' \emph{Ann. Neurol.}, vol.~63, pp. 193--203, Feb 2008.

\bibitem{pmid15792902}
L.~Astolfi, F.~Cincotti, D.~Mattia, C.~Babiloni, F.~Carducci, A.~Basilisco,
  P.~M. Rossini, S.~Salinari, L.~Ding, Y.~Ni, B.~He, and F.~Babiloni,
  ``{{A}ssessing cortical functional connectivity by linear inverse estimation
  and directed transfer function: simulations and application to real data},''
  \emph{Clin Neurophysiol}, vol. 116, pp. 920--932, Apr 2005.

\bibitem{icaconn}
L.~Astolfi, H.~Bakardjian, F.~Cincotti, D.~Mattia, M.~G. Marciani,
  F.~De~Vico~Fallani, A.~Colosimo, S.~Salinari, F.~Miwakeichi, Y.~Yamaguchi,
  P.~Martinez, A.~Cichocki, A.~Tocci, and F.~Babiloni, ``{{E}stimate of
  causality between independent cortical spatial patterns during movement
  volition in spinal cord injured patients},'' \emph{Brain Topogr}, vol.~19,
  pp. 107--123, 2007.

\bibitem{mvarica}
G.~G\'omez-Herrero, M.~Atienza, K.~Egiazarian, and J.~L. Cantero,
  ``{{M}easuring directional coupling between {E}{E}{G} sources},''
  \emph{NeuroImage}, vol.~43, pp. 497--508, Nov 2008.

\bibitem{cicaar}
M.~Dyrholm, S.~Makeig, and L.~K. Hansen, ``{{M}odel selection for convolutive
  {I}{C}{A} with an application to spatiotemporal analysis of {E}{E}{G}},''
  \emph{Neural Comput}, vol.~19, pp. 934--955, Apr 2007.

\bibitem{Valdes0501}
P.~A. Vald\'es-Sosa, J.~M. S\'anchez-Bornot, A.~Lage-Castellanos,
  M.~Vega-Hern\'andez, J.~Bosch-Bayard, L.~Melie-Garc\'ia, and
  E.~Canales-Rodr\'iguez, ``Estimating brain functional connectivity with
  sparse multivariate autoregression,'' \emph{Philosophical Transactions of the
  Royal Society B}, vol. 360, pp. 969--981, 2005.

\bibitem{Sanchez0801}
J.~M. S\'anchez-Bornot, E.~Mart\'inez-Montes, A.~Lage-Castellanos,
  M.~Vega-Hern\'andez, and P.~A. Vald\'es-Sosa, ``{Uncovering sparse brain
  effective connectivity: A voxel-based approach using penalized regression},''
  \emph{Statistica Sinica}, vol.~18, no.~4, 2008.

\bibitem{kernelgranger2}
D.~Marinazzo, M.~Pellicoro, and S.~Stramaglia, ``{{K}ernel method for nonlinear
  Granger Causality},'' \emph{Phys. Rev. Lett.}, vol. 100, p. 144103, 2008.

\bibitem{attias98}
H.~Attias and C.~E. Schreiner, ``{{B}lind source separation and deconvolution:
  the dynamic component analysis algorithm},'' \emph{Neural Comput}, vol.~10,
  pp. 1373--1424, Aug 1998.

\bibitem{parra00}
L.~Parra and C.~Spence, ``Convolutive blind source separation of non-stationary
  sources,,'' \emph{IEEE Trans. Speech Audio Processing}, vol.~8, no.~3, pp.
  320--327, 2000.

\bibitem{anemuller03}
J.~Anem{\"{u}}ller, T.~J. Sejnowski, and S.~Makeig, ``{{C}omplex independent
  component analysis of frequency-domain electroencephalographic data},''
  \emph{Neural Netw}, vol.~16, pp. 1311--1323, Nov 2003.

\bibitem{lBFGS}
\BIBentryALTinterwordspacing
J.~Nocedal, ``Updating quasi-newton matrices with limited storage,''
  \emph{Mathematics of Computation}, vol.~35, no. 151, pp. 773--782, 1980.
  [Online]. Available: \url{http://www.jstor.org/stable/2006193}
\BIBentrySTDinterwordspacing

\bibitem{Neal98}
R.~Neal and G.~E. Hinton, ``A view of the em algorithm that justifies
  incremental, sparse, and other variants,'' in \emph{Learning in Graphical
  Models}.\hskip 1em plus 0.5em minus 0.4em\relax Kluwer Academic Publishers,
  1998, pp. 355--368.

\bibitem{dal09}
R.~Tomioka and M.~Sugiyama, ``Dual augumented lagrangian method for efficient
  sparse reconstruction,'' \emph{IEEE Signal Proc Let}, vol.~16, no.~2, pp.
  1067--1070, 2009.

\bibitem{nolteleadfield}
G.~Nolte and G.~Dassios, ``Analytic expansion of the {EEG} lead field for
  realistic volume conductors,'' \emph{Phys. Med. Biol.}, vol.~50, pp.
  3807--3823, 2005.

\bibitem{montreal}
C.~J. Holmes, R.~Hoge, L.~Collins, R.~Woods, A.~Toga, and A.~C. Evans,
  ``Enhancement of {MR} images using registration for signal averaging,''
  \emph{J. Comput. Assist. Tomogr.}, vol.~22, no.~2, pp. 324--333, 1998.

\bibitem{molgedey94}
L.~Molgedey and H.~G. Schuster, ``{{S}eparation of a mixture of independent
  signals using time delayed correlations},'' \emph{Phys. Rev. Lett.}, vol.~72,
  pp. 3634--3637, Jun 1994.

\bibitem{belouchrani}
\BIBentryALTinterwordspacing
A.~Belouchrani, K.~Abed-Meraim, J.~F. Cardoso, and E.~Moulines, ``A blind
  source separation technique using second-order statistics,'' \emph{IEEE Trans
  Signal Proc}, vol.~45, no.~2, pp. 434--444, August 1997. [Online]. Available:
  \url{http://dx.doi.org/10.1109/78.554307}
\BIBentrySTDinterwordspacing

\bibitem{Ziehe98}
A.~Ziehe and K.-R. M{\"u}ller, ``{TDSEP--an efficient algorithm for blind
  separation using time structure},'' \emph{Proc. Int. Conf. on Artificial
  Neural Networks (ICANN '98)}, pp. 675--680, 1998.

\bibitem{ZieMueNolMacCur00}
A.~Ziehe, K.-R. M\"uller, G.~Nolte, and B.-M.~M. a~nd G.~Curio, ``Artifact
  reduction in magnetoneurography based on time-delayed second-order
  correlations,'' vol.~47, no.~1, pp. 75--87, January 2000.

\bibitem{ffdiag}
\BIBentryALTinterwordspacing
A.~Ziehe, P.~Laskov, G.~Nolte, and K.-R. M\"{u}ller, ``A fast algorithm for
  joint diagonalization with non-orthogonal transformations and its application
  to blind source separation,'' \emph{J. Mach. Learn. Res.}, vol.~5, pp.
  777--800, 2004. [Online]. Available:
  \url{http://portal.acm.org/citation.cfm?id=1016784}
\BIBentrySTDinterwordspacing

\bibitem{pairing04}
P.~Tichavsk\'y and Z.~Koldovsk\'y, ``Optimal pairing of signal components
  separated by blind techniques,'' \emph{IEEE Signal Proc Let}, vol.~11, no.~2,
  pp. 119--122, 2004.

\bibitem{BunMeiKirMue09}
P.~von B\"unau, F.~C. Meinecke, F.~Kiraly, and K.-R. M\"uller, ``Estimating the
  stationary subspace from superimposed signals,'' \emph{Physical Review
  Letters}, vol. 103, p. 214101, 2009.

\bibitem{neuroimage08}
S.~Haufe, V.~Nikulin, A.~Ziehe, K.-R. M\"uller, and G.~Nolte, ``Combining
  sparsity and rotational invariance in {EEG/MEG} source reconstruction,''
  \emph{NeuroImage}, vol.~42, no.~2, pp. 26–--738, 2008.

\bibitem{nips08}
S.~Haufe, V.~V. Nikulin, A.~Ziehe, K.-R. M\"uller, and G.~Nolte, ``Estimating
  vector fields using sparse basis field expansions,'' in \emph{Advances in
  Neural Information Processing Systems 21}, 2009.

\end{thebibliography}
 
\end{document}